\documentclass{aa}
\newbox\grsign \setbox\grsign=\hbox{$>$}
\newdimen\grdimen \grdimen=\ht\grsign
\newbox\laxbox \newbox\gaxbox
\setbox\gaxbox=\hbox{\raise.5ex\hbox{$>$}\llap
     {\lower.5ex\hbox{$\sim$}}}\ht1=\grdimen\dp1=0pt
\setbox\laxbox=\hbox{\raise.5ex\hbox{$<$}\llap
     {\lower.5ex\hbox{$\sim$}}}\ht2=\grdimen\dp2=0pt

\def\lax{\mathrel{\copy\laxbox}}
\def\chisqnu{$\chi^2_\nu$}

\begin{document}

\input psfig

\thesaurus{03         
          (13.07.1;   
           13.18.2;   
           13.09.2)}  

\title{SCUBA sub-millimeter observations of gamma-ray bursters}

\subtitle{I. GRB 970508, 971214, 980326, 980329, 980519, 980703}

\author{I.A. Smith\inst{1}
\and R.P.J. Tilanus\inst{2}
\and J. van Paradijs\inst{3,4}
\and T.J. Galama\inst{3}
\and P.J. Groot\inst{3}
\and P. Vreeswijk\inst{3}
\and L.B.F.M. Waters\inst{3}
\and C. Kouveliotou\inst{5}
\and R.A.M.J. Wijers\inst{6}
\and N. Tanvir\inst{7}}

\institute{Department of Space Physics and Astronomy,
Rice University, MS-108, 6100 South Main, Houston, TX 77005-1892 USA
\and Joint Astronomy Centre, 660 N. Aohoku Place, Hilo, HI 96720 USA
\and Astronomical Institute `Anton Pannekoek', 
University of Amsterdam and Center for High-Energy Astrophysics,\\
Kruislaan 403, 1098 SJ Amsterdam, The Netherlands
\and Department of Physics, University of Alabama in Huntsville,
Huntsville, AL 35899 USA
\and Universities Space Research Association,
NASA Marshall Space Flight Center, ES-62, Huntsville, AL 35812 USA
\and Department of Physics and Astronomy, SUNY,
Stony Brook, NY 11794-3800 USA
\and Institute of Astronomy, Cambridge University, Madingley Road,
Cambridge, CB3 0HA, UK}

\offprints{Ian A. Smith}

\date{Received ; accepted }

\maketitle

\begin{abstract}
We discuss the first results of our ongoing program of Target of Opportunity 
observations of gamma-ray bursts (GRBs) 
using the SCUBA instrument on the James Clerk Maxwell Telescope.
We present the results for GRB 970508, 971214, 980326,
980329, 980519, and 980703.

Our most important result to date is the detection of a fading 
counterpart to GRB 980329 at 850 $\mu$m.
Although it proved to be difficult to find the infrared counterpart to
this burst, the sub-millimeter flux was relatively bright.
This indicates that intrinsically the brightness of this counterpart was 
very similar to GRB 970508.
The radio through sub-millimeter spectrum of GRB 980329 is well fit by a
power law with index $\alpha = +0.9$.
However, we cannot exclude a $\nu^{1/3}$ power law attenuated by 
synchrotron self-absorption.
An $\alpha \sim +1$ VLA-SCUBA power law spectrum is definitely ruled
out for GRB 980703, and possibly also for GRB 980519.

We cannot rule out that part of the sub-millimeter flux from GRB 980329 
comes from a dusty star-forming galaxy at high redshift, such as the ones 
recently discovered by SCUBA.
Any quiescent dust contribution will be much larger at sub-millimeter than
at radio wavelengths.
Both a high redshift and large dust extinction would help explain 
the reddening of the counterpart to GRB 980329, and a redshift of
$z \sim 5$ has been suggested.
The large intensity of this burst might then indicate
that beaming is important.


\end{abstract}

\section{Introduction}

An important reason for the long-standing mysteries surrounding
the gamma-ray burst (GRB) sources has been the lack of prompt
accurate locations to look for quiescent or fading counterparts.
This situation changed dramatically in 1997 with the rapid and
accurate location of X-ray emission detected during GRBs using the 
wide-field camera on the Satellite per Astronomia X (BeppoSAX),
the All Sky Monitor on the {\it Rossi X-ray Timing Explorer} ({\it RXTE}), 
and from the detection of X-ray afterglow in scans of BATSE GRB error boxes
with the PCA on {\it RXTE}.

These localized X-ray counterparts have led to intense multiwavelength
campaigns.
Optical transients have been found to some but not all of the bursts with 
fading X-ray counterparts.
The reason optical transients have not been discovered for the other
bursts with X-ray counterparts (some of which are much brighter during
the gamma-ray burst itself) is that the searches have not always been
sensitive enough, particularly given the problem of absorption local to 
the source (Groot et al. \cite{scuba:ggvp98}).

In spite of detailed searches, radio emission has only been detected for 
a few bursts.
The first detection of a variable radio source was GRB 970508
(Frail et al. \cite{scuba:fra97}; Taylor et al. \cite{scuba:tay97};
Galama et al. \cite{scuba:galama98a}).
This $\sim$ mJy source had a {\it rising} spectral index, and its rapid radio
variability was most likely caused by interstellar scintillation.
The combined observations of GRB 970508 showed that the peak of the
spectrum was in the sub-millimeter region 
(Galama et al. \cite{scuba:galama98b}) with the radio,
millimeter, and optical emission peaking days to weeks after the burst
(Frail et al. \cite{scuba:fra97}; Galama et al. \cite{scuba:galama98a};
Gruendl et al. \cite{scuba:gru98};
Bremer et al. \cite{scuba:bre98}; Pian et al. \cite{scuba:pian98};
Pedersen et al. \cite{scuba:ped98}; Castro-Tirado et al. \cite{scuba:cas98}).
Even before any counterparts were found, two completely separate
classes of models had suggested this would be the case:
(1) cosmological fireball models 
(e.g. Paczy\'nski \& Rhoads \cite{scuba:pac93}; Katz \cite{scuba:katz94};
M\'esz\'aros \& Rees \cite{scuba:mes97}), and
(2) Compton scattering models (e.g. Liang et al. \cite{scuba:lia97etal};
Liang \cite{scuba:lia97}).

For GRB 970508 there are at least two breaks in the spectrum between
$10^{11}$ and $10^{14}$ Hz (Galama et al. \cite{scuba:galama98b}) 
making this a crucial region for the models.
To obtain a complete picture of the nature of the burst counterparts, it
is clear that one needs to cover the entire spectrum, and sub-millimeter
observations with a $\sim$ mJy sensitivity are needed.
This is particularly important since the optical emission can be suppressed 
by local absorption, and the radio emission can be self absorbed as well as
scrambled by interstellar scintillation (Walker \cite{scuba:wal98}), while 
``clean'' observations ought to be possible at sub-millimeter wavelengths.

In this paper we discuss our ongoing program of Target of Opportunity 
observations using SCUBA on the James Clerk Maxwell Telescope.
In \S 2 we discuss some of the technical features of SCUBA that make it
well suited for performing counterpart searches.
In \S 3 we present the results of our observations to date on
GRBs 970508 (which was a limited trial run), 971214, 980326, 980329,
980519, and 980703.
In \S 4 we give a brief discussion.

\section{SCUBA Details}

SCUBA is the new sub-millimeter continuum instrument for the James Clerk 
Maxwell Telescope on Mauna Kea, Hawaii (for a review see 
Holland et al. \cite{scuba:hol98}).
It uses two arrays of bolometers to simultaneously observe the same
region of sky, $\sim 2.3\arcmin$ in diameter.
The arrays are optimized for operations at 850 and 450 $\mu$m.
Fully sampled maps of the $2.3\arcmin$ region can be made by ``jiggling'' 
the array.
Thus this mode is appropriate for mapping the better localized GRB error
boxes as well as those of the X-ray transients.
This mode can also be used to look for extended quiescent counterparts.

Deeper photometry can be performed
using just the central pixel of these arrays.
There are also dedicated photometry pixels for 1100, 1350, and 2000 
$\mu$m observations (that cannot be used at the same time as the arrays).
This photometry mode is appropriate for well localized radio 
or optical transients.

Scan mapping of larger GRB error boxes using the 450:850 filters 
is also possible.
While this mode is ideal for mapping the long thin GRB error boxes that are
obtained using triangulation between satellites, e.g. 
$5\arcmin \times 0.5\arcmin$ (Hurley et al. \cite{scuba:hur97}), the
sensitivity is greatly reduced.

Given the rapid dissemination of candidate optical and radio transients, 
the photometry mode is the one that we use most often.
In principle, the most sensitive measurements can be made at 1350 $\mu$m, 
though the 850 $\mu$m array has an advantage because the multiple bolometers
permit a good sky noise subtraction.
In photometry mode, for an integration time of 2 hours, we would expect 
to achieve an rms $\sim 1$ mJy at 1350 $\mu$m, $\sim 1.5$ mJy
at 850 $\mu$m, and $\sim 5 - 20$ mJy at 450 $\mu$m. 
The sensitivities depend significantly on the weather, particularly 
at the shorter wavelengths.
The jiggle maps give an rms that is a factor $\sim 2-3$ times higher.

\section{Results of SCUBA Observations}

\subsection{GRB 970508}

As a limited trial run, a 30 minute SCUBA observation of GRB 970508 was 
made on 1997 May 26 using the 1350 $\mu$m photometry pixel.
The weather conditions were very poor.
No source was detected with an rms $\sim 10$ mJy.
This result is consistent with the other millimeter observations of GRB 970508
(Gruendl et al. \cite{scuba:gru98}; Bremer et al. \cite{scuba:bre98};
Shepherd et al. \cite{scuba:she98}).

\subsection{GRB 971214}

Our preliminary SCUBA results on GRB 971214 were originally reported in 
Smith et al. (\cite{scuba:smi97}).

The BeppoSAX GRB Monitor was triggered on 1997 December 14.97 UT
(Heise et al. \cite{scuba:heise97}).
A previously unknown fading X-ray source (1SAX J1156.4+6513) was found 
inside the burst error circle (Antonelli et al. \cite{scuba:ant97}).
Consistent with this X-ray source an optical transient was found 
(e.g. Halpern et al. \cite{scuba:halp98}; 
Kulkarni et al. \cite{scuba:kul98};
Ramaprakash et al. \cite{scuba:ram98};
Gorosabel et al. \cite{scuba:gor98};
Diercks et al. \cite{scuba:die98}).
A possible quiescent host to the transient was found with a redshift $z=3.42$
(Kulkarni et al. \cite{scuba:kul98}).
No radio counterpart has so far been seen 
(Ramaprakash et al. \cite{scuba:ram98}).

\begin{figure}[t]
\vspace{0cm}
\hspace{0cm}\psfig{figure=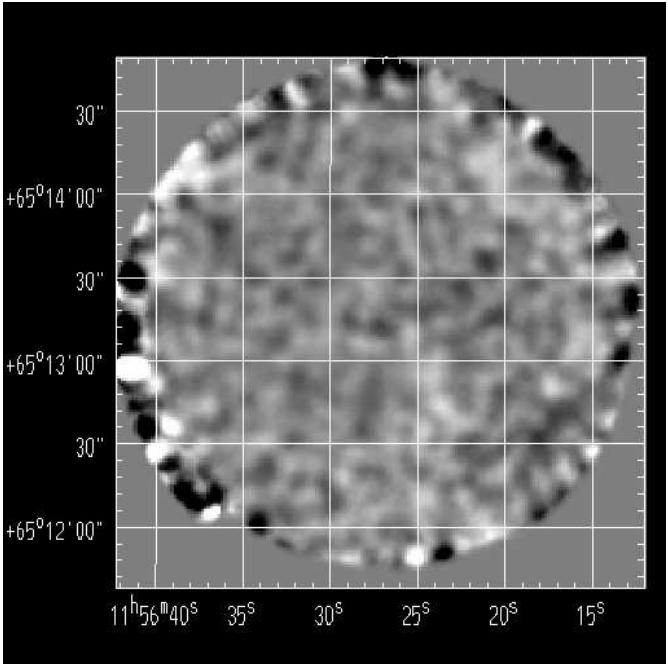,width=8.8cm}
\vspace{0cm}
\caption[]{SCUBA 850 $\mu$m jiggle map of the $\sim 1\arcmin$ radius error 
box to 1SAX J1156.4+6513 on UT 1997 December 16.
The beam size is $14.7\arcsec$.}
\label{figure1}
\end{figure}

We began our series of SCUBA observations on UT 1997 December 16,
before the optical transient was reported, and when the error box of 
1SAX J1156.4+6513 had a radius $\sim 1\arcmin$
(L. Piro, private communication).
We performed a 450:850 jiggle map of the whole error box.
Fig.~\ref{figure1} shows the whole 850 $\mu$m map, which illustrates a
typical SCUBA jiggle map.
At 850 $\mu$m, the rms was 3 mJy in the central region of the map and
the beam size was $14.7\arcsec$.
The optical transient was near the edge of this map at
RA(J2000) = 11:56:26.4, DEC(J2000) = +65:12:00.5,
where the rms is approximately 5 mJy.  
No sources were detected by SCUBA anywhere in the map.

The remainder of our SCUBA observations were performed on the optical
transient using the photometric mode with the 450:850 arrays.
The results are given in Table \ref{table1}.
We did not detect a sub-millimeter continuum source at the location
of the optical transient.
Combining all our photometric observations gives an rms of 1.0 mJy
at 850 $\mu$m.

The decay of the optical flux followed a power law 
$\propto t^{-\delta}$ with slope $\delta \sim 1.2$.
In the simple adiabatic piston model, a fireball produced by a 
one time impulsive injection of energy in which only the forward blast
wave efficiently accelerates particles predicts a power law spectrum 
$S_\nu \propto \nu^{-\beta}$ with
energy spectral index $\beta = 2 \delta / 3$
(Wijers, Rees, \& M\'esz\'aros \cite{scuba:wij97}).
For GRB 971214, this would imply $\beta = 0.8$.
This is in general agreement with the optical to X-ray slope.
Extrapolating this power law gives a flux density $\lax 1$ mJy at
$850 \mu$m on December 17.
However, the optical spectrum alone has a much steeper spectrum, indicating
that there is significant extinction local to the source 
(Halpern et al. \cite{scuba:halp98}; Ramaprakash et al. \cite{scuba:ram98}).
The required correction is several magnitudes in the I-band, which similarly
raises the prediction at $850 \mu$m.
Our SCUBA limits then imply that there is a break between the optical and
sub-millimeter bands, as was found in GRB 970508.
Infrared observations of GRB 971214 suggest that this break was at
$\sim 1 ~\mu$m in the first few hours after the burst
(Ramaprakash et al. \cite{scuba:ram98}; Gorosabel et al. \cite{scuba:gor98}).

\begin{table}[t]
\caption[]{SCUBA observations of the optical transient to GRB 971214.
There is a $\sim 10\%$ systematic uncertainty in all the flux densities.
}
\label{table1}
\begin{flushleft}
\[
\begin{tabular}{ll}
\hline
\noalign{\smallskip}
Observing Time &  850 $\mu$m rms \\
(UT 1997) & (mJy) \\
\noalign{\smallskip}
\hline
\noalign{\smallskip}
Dec 16 & 5.0 \\
Dec 17 & 1.4 \\
Dec 19 & 1.9 \\
Dec 22 & 1.3 \\
\noalign{\smallskip}
\hline
\end{tabular}
\]
\end{flushleft}
\end{table}

\subsection{GRB 980326}

The BeppoSAX GRB Monitor was triggered on 1998 March 26.888 UT
(Celidonio et al. \cite{scuba:cel98}).
Although BeppoSAX was unable to make an observation with the Narrow Field
Instruments and {\it RXTE} did not see any X-ray emission from the GRB
error box (Marshall \& Takeshima \cite{scuba:mar98}), a candidate
optical transient was found (Groot et al. \cite{scuba:gro98b}).
This transient was notable for its unusually rapid optical fade, with
a power law decay index $\delta = 2.10 \pm 0.13$ 
(Eichelberger et al. \cite{scuba:eic98}; Groot et al. \cite{scuba:gro98b}).
A constant underlying source with $R_c = 25.5 \pm 0.5$ was found 
(Grossan et al. \cite{scuba:gros98}; Djorgovski et al. \cite{scuba:djo98a};
Groot et al. \cite{scuba:gro98b}).
This burst was also interesting in that the gamma-ray spectrum during the
burst was quite soft.

We used SCUBA to make a short photometry observation of the optical transient 
to GRB 980326 on 1998 March 29.
The source was not detected, with rms 2.5 mJy at 850 $\mu$m and 25 mJy at
450 $\mu$m.
Since there was no report of a radio counterpart, and the much more 
interesting GRB 980329 occurred at this time, we did not try to make any 
further observations of GRB 980326.

On 1998 March 29 the counterpart had $R_c = 24.5$, and the optical
spectrum was poorly determined, with $\beta = 0.66 \pm 0.70$ 
(Groot et al. \cite{scuba:gro98b}).
Extrapolating with $\beta = 0.66$ would give a flux density of 
0.06 mJy at 850 $\mu$m, while using the $1 \sigma$ value of 
$\beta = 1.36$ gives 8.8 mJy at 850 $\mu$m.
Possible distortions to the optical spectrum from the underlying quiescent
source, and the difficulty in determining the extinction corrections add 
to the uncertainty in the counterpart spectrum.
Thus we are currently unable to make any statements about breaks in the 
optical to sub-millimeter spectrum for GRB 980326.

\subsection{GRB 980329}

Our preliminary SCUBA results on GRB 980329 were originally reported in 
Smith \& Tilanus (\cite{scuba:st98}).

\begin{figure}[t]
\vspace{0cm}
\hspace{0cm}\psfig{figure=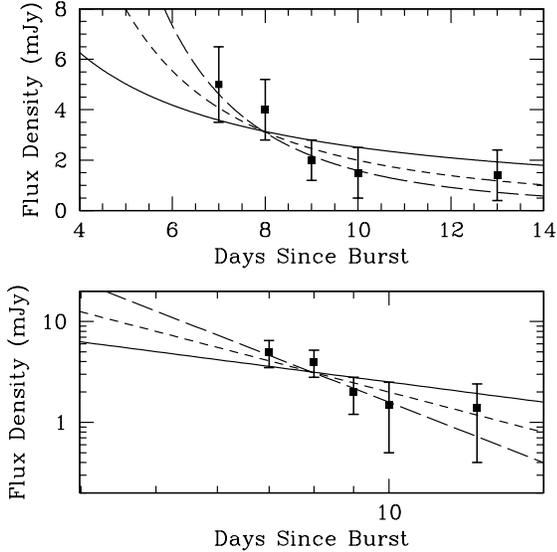,width=8.8cm}
\vspace{0cm}
\caption[]{Evolution of the 850 $\mu$m flux seen by SCUBA for the 
counterpart VLA J070238.0+385044 to GRB 980329.
The two plots show the same data plotted on linear-linear and log-log axes.
The curves plot flux density $\propto t^{-m}$ where $t$ is the time since
the burst and $m=1$ (solid), $m=2$ (short dashed), and $m=3$ (long dashed).}
\label{figure2}
\end{figure}

The BeppoSAX GRB Monitor was triggered on 1998 March 29.156 UT
(Frontera et al. \cite{scuba:front98}).
This was the brightest burst that had been seen simultaneously by the BeppoSAX 
Wide Field Camera, with a peak flux $\sim 6$ Crab in the 2--26 keV band.
A fading X-ray source 1SAX J0702.6+3850 was found using the BeppoSAX 
Narrow Field Instruments (in't Zand et al. \cite{scuba:intz98}).
Inside this X-ray error box, a variable radio source VLA J070238.0+385044
was found that was similar to GRB 970508
(Taylor et al. \cite{scuba:tay98a,scuba:tay98b}).
It was not until after the variable radio source was discovered that infrared 
observations found a fading counterpart
(Klose et al. \cite{scuba:klose98}; Palazzi et al. \cite{scuba:pal98};
Metzger \cite{scuba:metz98}): 
this indicated that the optical extinction was significant for this source
(Larkin et al. \cite{scuba:lar98}; Taylor et al. \cite{scuba:tay98b}).
A possible host galaxy was found at this location
(Djorgovski et al. \cite{scuba:djo98b}).

Starting on 1998 April 5, we made a series of observations of 
VLA J070238.0+385044 using SCUBA.
%
On April 5.2 UT, we detected the source at 850 $\mu$m with a flux density 
of $5 \pm 1.5$ mJy.
This source was confirmed on April 6.2 with a flux density of $4 \pm 1.2$ mJy, 
resulting in an average of $4.5 \pm 1$ mJy over the two days.
The source was not detected at 450 $\mu$m, with an rms of 10.0 mJy averaged 
over these two days.  
The 850 $\mu$m source was present in all our separate integrations, making us
confident that it was real.  
A hint of a fading trend was confirmed by observations on April 7.2, when 
the 850 $\mu$m flux density was $2 \pm 0.8$ mJy.  
Observations on April 8 gave $1.5 \pm 1.0$ mJy at 850 $\mu$m,
with no detection at 1350 $\mu$m (the rms was 1.2 mJy).
Finally, the signal was $1.4 \pm 1.0$ mJy at 850 $\mu$m on April 11.

Assuming the sub-millimeter fluxes are due to the burst
counterpart, they should represent ``clean'' measures of its
intensity, unaffected by scintillation and extinction.
Although the optical emission was significantly reduced in this burst,
the radio and sub-millimeter observations show that intrinsically the
brightness of this counterpart was similar to GRB 970508 (e.g. see Figure 2
of Palazzi et al. \cite{scuba:pal98}).

\begin{figure}[t]
\vspace{0cm}
\hspace{0cm}\psfig{figure=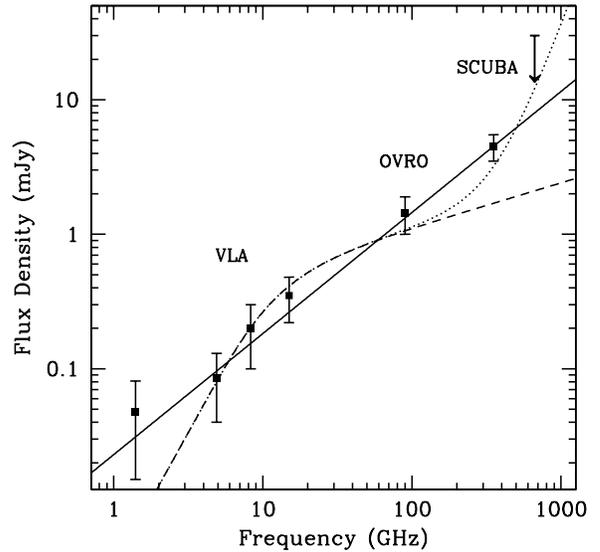,width=8.8cm}
\vspace{0cm}
\caption[]{The radio through sub-millimeter spectrum of GRB 980329.
The time-averaged VLA and OVRO points are taken from Figure 2 of
Taylor et al. (\cite{scuba:tay98b}).
The $3 \sigma$ SCUBA upper limit at 450 $\mu$m has been included.
The solid curve is a power law $S_\nu \propto \nu^\alpha$ with
$\alpha = + 0.9$.
The dashed curve is a $\nu^{1/3}$ power law attenuated by a synchrotron
self-absorption with $\nu_0 = 13$ GHz at $\tau_0 = 1$.
The dotted curve adds a $\nu^3$ power law to the synchrotron curve.}
\label{figure3}
\end{figure}

Fig.~\ref{figure2} plots the evolution of the 850 $\mu$m SCUBA flux.
For a power law decay with the flux density $\propto t^{-m}$ where $t$ is the 
time since the burst, the best fit power law index is $m = 2.8$.
However, $m$ is not tightly constrained: the 90\% confidence interval is
$m = 0.9$ to $m = 5.6$.

Fig.~\ref{figure3} adds the SCUBA results to the VLA-OVRO results presented in
Figure 2 of Taylor et al. (\cite{scuba:tay98b}).
Because of the averaging of the rapidly varying radio data over several days,
some caution is required in using this figure.
Taylor et al. found that a power law $S_\nu \propto \nu^\alpha$ with
$\alpha = + 0.9$ gave the best fit to the VLA-OVRO data alone.
The solid curve shows that this extends very well to the SCUBA results.
The dashed curve shows that the popular $\nu^{1/3}$ power law
(e.g. Katz \cite{scuba:katz94}; Waxman \cite{scuba:wax97}) 
attenuated by a synchrotron self-absorption component gives a much worse
description of the longer wavelength emission for GRB 980329.
However, the reduced \chisqnu = 2.6 for this fit, and the probability that 
a random set of data points would give a value of \chisqnu\ as large or 
larger than this is $Q = 0.034$.
It is therefore not possible to exclude this model, and it will be important
to study more bursts to determine whether there is a range of power law 
indices for $\alpha$.

One way to slightly reduce the sub-millimeter flux of the counterpart in 
Fig.~\ref{figure3} would be if part of the flux comes from an underlying
quiescent sub-millimeter source.
There is a suggestion in Fig.~\ref{figure2} that the 850 $\mu$m flux is
leveling off, but an instrument more sensitive than SCUBA will be required 
to see if such a quiescent source is present for GRB 980329.
SCUBA has recently discovered several dusty star-forming galaxies at high 
redshifts (Hughes et al. \cite{scuba:hughes98}; 
Barger et al. \cite{scuba:barger98}; Smail et al. \cite{scuba:smail98}).
Both a high redshift and large dust extinction would help explain 
the reddening of the counterpart to GRB 980329, and a redshift of 
$z \sim 5$ has been suggested (Fruchter \cite{scuba:fruct98}).
The large intensity of this burst might then indicate that beaming 
is important.

In dust models, one expects $S_\nu \propto \nu^3$ or $\nu^4$
(e.g. Dwek \& Werner \cite{scuba:dwek81}).
Thus any quiescent dust contribution is very much larger at sub-millimeter 
than at radio wavelengths.
For illustrative purposes the dotted curve in Fig.~\ref{figure3} adds a 
quiescent $S_\nu \propto \nu^3$ component to the synchrotron curve.
In this example, the quiescent flux density at 8.3 GHz is only 0.02 $\mu$Jy.

\subsection{GRB 980519}

The BeppoSAX GRB Monitor was triggered on 1998 May 19.514 UT
(Muller et al. \cite{scuba:mul98}).
A fading X-ray counterpart 1SAX J2322.3+7716 was found, although the X-ray 
decay was not monotonic (Nicastro et al. \cite{scuba:nic98}).
A fading optical counterpart was also found, whose power law decay was
steep $\delta \sim 2$ (e.g. Jaunsen et al. \cite{scuba:jau98}, 
Djorgovski et al. \cite{scuba:djo98c}).
A very faint quiescent optical source was eventually detected
(Sokolov et al. \cite{scuba:sok98}, Bloom et al. \cite{scuba:bloom98a}).
A variable radio source was found at the same location
(Frail et al. \cite{scuba:fra98a}).

This source was not in an ideal location for SCUBA observations, with 
the elevation never rising above $35\degr$.
Also, the weather conditions were very poor at this time, and the JCMT was
locked into using a different instrument.
This meant we were only able to make one SCUBA observation of GRB 980519
on 1998 May 27
The source was not detected, with flux density $0.9 \pm 1.8$ mJy at 850 
$\mu$m and an rms of 80 mJy at 450 $\mu$m.

The radio flux uncorrected for scintillation at the time of our SCUBA 
observation is not currently available.
On May 22.3, the 8.3 GHz flux measured by the VLA was 0.1 mJy.
Extrapolating from this using $\alpha = +1/3$ predicts a flux density 
of 0.35 mJy at 850 $\mu$m.
On the other hand, extrapolating using $\alpha = +1$ predicts a flux
density of 4.3 mJy at 850 $\mu$m.
When the final radio results are available, it may be possible to determine 
whether this steeper slope is unacceptable for GRB 980519.

It is believed that the optical extinction is small for this burst
(Gal et al. \cite{scuba:gal98}).
Assuming the optical flux continued to decay with a power law of index
$\delta = 1.98$, and extrapolating the optical spectrum assuming a power 
law index of $\beta = 1.26$ (Gal et al. \cite{scuba:gal98}) would predict 
a flux of 1.6 mJy at 850 $\mu$m at the time of our SCUBA observation.
Unfortunately, our observation was made too late to determine if there
was a break between the optical and millimeter bands in GRB 980519.

\subsection{GRB 980703}

BATSE trigger 6891 (Kippen et al. \cite{scuba:kip98})
was also detected by the {\it RXTE} ASM on
1998 July 3.182 UT (Levine et al. \cite{scuba:lev98}).
BeppoSAX NFI observations of the {\it RXTE} ASM error box located
a fading X-ray source 1SAX J2359.1+0835
(Galama et al. \cite{scuba:galama98c,scuba:galama98d}).
A variable radio, infrared, and optical counterpart was found, as well as
an underlying galaxy with $R = 22.6$ and a redshift of 0.966
(e.g. Bloom et al. \cite{scuba:bloom98b};
Djorgovski et al. \cite{scuba:djo98d}).

SCUBA observed the radio counterpart on 1998 July 10.5 UT.
The source was not detected, with an rms of 2.7 mJy at 1350 $\mu$m.
A second observation was performed on 1998 July 15.6 UT.
Again the source was not detected with an rms of 2.1 mJy at 850 $\mu$m
and 24 mJy at 450 $\mu$m.
Another observation was tried on July 16, but the weather conditions were
too poor to produce any useful results.

While the 4.86 GHz flux suffered from large variations, the 8.46 GHz flux
was steadier, with a mean of 0.94 mJy (Frail et al. \cite{scuba:fra98b}).
Extrapolating from this using $\alpha = +1$ predicts flux densities 
of 39 and 25 mJy at 850 and 1350 $\mu$m respectively.
Our SCUBA results can definitely rule out this simple power law 
for GRB 980703.
Extrapolating from the radio using $\alpha = +1/3$ predicts flux densities
of 3.3 and 2.8 mJy at 850 and 1350 $\mu$m respectively.
While we would expect to have detected $1 - 2 \sigma$ signals in our SCUBA
observations, the lack of detections are not inconsistent with this model
for GRB 980703.

\section{Discussion}

The sub-millimeter is an important band for GRB studies because it is where
the emission peaks in some bursts in the days to weeks following the burst.
The sub-millimeter emission is not affected by extinction local to the source
or interstellar scintillation.
We have shown that sub-millimeter observations are important to:

\begin{itemize}

\item
Determine the radio to sub-millimeter spectrum, and see if it agrees
with the synchrotron models.

\item
Discover whether there are breaks in the optical to sub-millimeter spectrum.

\item
Determine the evolution of the sub-millimeter flux.

\item
Look for underlying quiescent sources that may be dusty star-forming 
galaxies at high redshifts.

\end{itemize}

To obtain a detailed understanding of the GRB counterpart behaviors will
require observations of many bursters.
To this end our program of Target of Opportunity observations using 
SCUBA is ongoing.

\begin{acknowledgements}

The James Clerk Maxwell Telescope is operated by The Joint Astronomy 
Centre on behalf of the Particle Physics and Astronomy Research Council of the 
United Kingdom, the Netherlands Organisation for Scientific Research, and the
National Research Council of Canada.

We thank the JCMT Director Ian Robson for authorizing the ToO observations,
and Iain Coulson, Fred Baas, Wayne Holland, and Graeme Watt for their valuable
assistance with the observations and reductions.
We thank Lorne Avery and John MacLeod for their cooperation and
valuable help in obtaining the observations of GRB 980329.

We are grateful to the BeppoSAX team, to Scott Barthelmy and Paul Butterworth
of The GRB Coordinates Network (GCN), and to the other ground-based observers 
for the rapid dissemination of their burst results.

This work was supported by NASA grant NAG 5-3824 at Rice University.

\end{acknowledgements}

\end{document}